# Competitive Adsorption in Polymer Nanocomposites: The Molecular Weight and End-Group Effect Revealed by SANS and MD Simulations

Tae Yeon Kong[1], WooJin Kim[2], YongJoo Kim[3,*], So Youn Kim[1,*]

[1]Department of Chemical and Biological Engineering, Institute of Chemical Processes, Seoul National University; Seoul, 08826, Republic of Korea

[2]Department of Materials Science and Engineering, Kookmin University; Seoul 02707, Republic of Korea

[3]Department of Materials Science and Engineering, Korea University; Seoul, 02841, Republic of Korea

## ABSTRACT

Understanding polymer adsorption at interfaces is essential for designing advanced polymer-based nanomaterials with tailored interfacial properties. Although adsorption significantly influences the macroscopic properties of polymer composites and thin films, a comprehensive understanding of molecular weight (MW)-dependent adsorption remains challenging and controversial, particularly in polydisperse polymer systems, due to the limitations of experimental approaches.

We investigate competitive adsorption in bidisperse poly(ethylene glycol) (PEG) melts and find that shorter chains preferentially adsorb onto nanoparticle surfaces. Experiments and molecular dynamics simulations reveal that the high density of terminal hydroxyl groups in short PEG chains strengthens hydrogen bonding at the interface, driving enthalpy-driven adsorption despite identical polymer backbones. This leads to a densely packed interfacial layer that alters the conformation of longer chains. These findings highlight the critical role of end-group functionality in interfacial polymer behavior and provide new insights for tailoring nanocomposite properties.

## INTRODUCTION

The adsorption of polymers at interfaces is a ubiquitous phenomenon, and the unique behavior of adsorbed polymers distinct from bulk polymers has garnered significant interest over the past few decades.[1-4] This interest stems from the substantial correlation between the behavior of adsorbed polymers and the macroscopic properties of materials, such as glass transition temperature,[5,6] viscosity,[7] thermal expansion,[8,9] and crystallization.[10] For example, composite materials are made up of polymers with filler particles, and the physical properties of such composites are determined by the interaction between the particle surfaces and the polymer chains.[11-15]

When polymers adsorb onto surfaces, they lose conformational freedom, leading to an entropy decrease and the degree of entropy reduction generally depends on the chain length.[16-19] For chemically identical polymer melts without any molecular weight (MW)-dependent energetic preference for the surface, shorter and stiffer chains are theoretically predicted to favor the adsorption because the entropy loss associated with the longer chains is greater due to their segmental connectivity.[19-21] Numerous theoretical studies, including density functional theory (DFT)[22,23] and self-consistent field theory (SCFT)[24,25], have reported that entropy-driven surface enrichment of shorter chains occurs in athermal bidisperse polymer melt. Jacob et al.[26] observed the surface enrichment of shorter chains using surface layer matrix-assisted laser desorption ionization time-of-flight mass spectrometry.

However, theoretical and experimental results have also been reported that longer chains are preferentially adsorbed onto the surface,[27-31] demonstrating that confirming the MW-dependent chain preference for adsorption remains challenging. Monte Carlo (MC) simulations employing the wall polymer reference interaction site model (wall-PRISM) for binary polymer mixtures consisting of flexible and stiff chains showed that stiff longer chains preferentially partition to neutral surfaces, primarily due to their superior local packing efficiency near the interface.[32] Experimental studies also indicated that pre-adsorbed short-chain polymers can be displaced by long-chain polymer solution. For example, Fu et al.[29,30] utilized coumarin end-labeled poly(ethylene oxide) (PEO) and demonstrated via total internal reflectance fluorescence that pre-adsorbed short-chain PEO was replaced by long-chain PEO, attributed to the increased number of segment-surface contacts achievable by longer chains. Similarly, Dijt et al.[31] observed through adsorption isotherm that short PEO chains adsorbed onto silica were displaced by longer ones within minutes attributed to the translational entropy gain.

Competitive adsorption in solution conditions is further influenced by solvent-induced interactions and the translational entropy compared to the melts, complicating the adsorption behavior. In dilute solutions, a part of the adsorbed chain can still interact with solvents, thereby reducing the configurational entropy penalty for longer chains.[33] Consequently, longer polymers may be preferentially adsorbed. Furthermore, the translational entropic gain per monomer is greater when more short chains are presented in the dilute solution.[31,34] However, as the polymer concentration increases, the entropic gain from the solvent interactions becomes progressively restricted and the impact of translational entropy is markedly reduced, causing this trend to reverse above a critical concentration.[35] As a result, a fair comparison without solvent contribution for competitive adsorption between different MW has not been realized and thus remains insufficiently understood requiring further experimental investigation.

To address this gap, we designed the bidisperse polymer nanocomposites (PNCs) to investigate MW-dependent competitive adsorption of poly(ethylene glycol) (PEG) melts, where polymer adsorption onto nanoparticles (NPs). Specifically, we first prepared PNCs by selectively adsorbing either long (or short) PEG onto nanoparticle surfaces. Subsequently, complementary short (or long) PEG were introduced, followed by thermal annealing under melt conditions to systematically examine competitive adsorption without solvent-induced interactions. We report the preferential adsorption of shorter chains, through direct experimental observations with small angle neutron scattering (SANS) and Fourier transform infrared (FT–IR) spectroscopy and molecular dynamics (MD) simulations. Although this outcome resembles previously reported results, our findings indicate that enthalpic contributions originating from end-groups play a more significant role than entropic effects. These results provide valuable insights into understanding and designing polymer interfaces in real systems, where complete elimination of enthalpic biases is practically unattainable.

## EXPERIMENTAL SECTION

### Preparation of Polymer Nanocomposite (PNC)

PNCs were prepared with poly(ethylene glycol) (PEG) and silica nanoparticles (NPs). PEG with different MWs, 0.4, 10 kg/mol, were purchased from Sigma Aldrich and used. Silica NPs with a diameter of 38 nm were synthesized by the Stöber method[36] and used as dispersed in ethanol. The weight fraction of the NPs solution was about 10 wt%. The average size of silica NPs was determined by DLS and by fitting SAXS intensity as a spherical form factor.

To investigate the competitive adsorption behavior in a bidisperse polymer melt, a proper amount of the silica NP solution ($\phi_c$=0.1 for final PNCs) was added to either 0.4k (short) or 10k (long) PEG, ensuring that the polymer of a particular MW predominantly adsorbed onto the particle surface, as illustrated in Figure S1. The mixed solutions were dried within approximately 2 hours to fully evaporate the solvent, forming the initial PNCs. These initial PNCs were then further annealed at 70 °C for 24 hours in a vacuum oven to obtain PNCs pre-adsorbed with either short or long polymer chains.

Subsequently, a polymer solution with a different MW from that of the pre-adsorbed chains (either 0.4k or 10k PEG) was added, followed by drying to prepare the final bidisperse PNCs with selectively pre-adsorbed chains.

To ensure sufficient chain mobility for chain exchange, the final PNCs were thermally annealed at 70 °C under vacuum conditions for 7 days. After this process, both thermally annealed and non-annealed samples were characterized to compare the modification of interfacial layers.

**Small Angle Neutron Scattering (SANS)**

SANS experiments were performed using the 40 m SANS instrument at HANARO of the Korean Atomic Energy Research Institute (KAERI) in Daejeon, Republic of Korea. Two sample–to–detector distances of 1.16 and 17.5 m were used to cover the q range of $0.004 < q < 0.4$ Å$^{-1}$. The scattered intensity was corrected for background, empty cell scattering, and the sensitivity of individual detector pixels. The corrected data sets were placed on an absolute scale through the secondary standard method. The samples were loaded into a 1 mm path-length quartz cell. The cell temperature was maintained at 70 °C.

**Determination of Partial Structure Factors from SANS**

The intensity, $I(q)$, of scattered neutrons at wave vector $q$, has three contributions:

$$I(q) \sim A\Delta\rho_c^2 P_c(q)S_{cc}(q) + B\Delta\rho_c\Delta\rho_p P_c(q)^{0.5}S_{pc}(q) + C\Delta\rho_p^2 S_{pp}(q)$$

Where $S_{ij}(q)$ are the structure factors associated with the two components ($pp$, $pc$, $cc$) where the subscripts $p$ and $c$ indicate polymer segments and particles, respectively; $\Delta\rho_j$ is the difference between the scattering length density (SLD, $\rho$) of component $j$ and the medium; $A$, $B$, and $C$ are constants.

To achieve the contrast matching method, SLD of nanoparticles, polymer segments, and solvent in highly concentrated polymer/particle solutions must be determined. While the cross sections of silica nanoparticles and PEG are fixed, the contrast relative to the background can be tuned by varying the D/H–EtOH ratio in the solvent phase as shown in Figure S2. In concentrated polymer solutions, the solvent cross-section reflects its composition and is written: $\rho_s = \phi_h\rho_h + \phi_d\rho_d + \phi_p\rho_p$, where $\rho_h$ and $\rho_d$ are the cross sections of H–EtOH and D$_6$–EtOH, respectively, while $\rho_p$ is the polymer scattering cross-section. Here, $\phi_h$, $\phi_d$ and $\phi_p$ represent the mass fractions of H–EtOH, D$_6$–EtOH, and polymer in the continuous phase, respectively, such that $\phi_h + \phi_d + \phi_p = 1$. We fixed $\phi_p = 0.4$ ($R_p = 0.4$) and varied $\phi_d/(\phi_h + \phi_d)$. Our initial studies established $\rho_c = 3.74 \times 10^{-6}$ Å$^{-2}$, $\rho_p = 6.60 \times 10^{-7}$ Å$^{-2}$, $\rho_h = -0.35 \times 10^{-6}$ Å$^{-2}$, $\rho_d = 6.16 \times 10^{-6}$ Å$^{-2}$ such that the contrast match condition for silica is achieved at $\phi_d/(\phi_h + \phi_d) \sim 0.96$ and for PEG at $\phi_d/(\phi_h + \phi_d) \sim 0.15$. Under condition $\phi_d/(\phi_h + \phi_d) \sim 0.96$, the measured scattering will be dominated by scattering from the polymer where contributions from $S_{pp}(q)$ and $S_{pc}(q)$ will be present since $\Delta\rho_c \sim 0$ in the above equation.

Scattering measurements at a fixed $\phi_c$ were made at five D/H–EtOH ratios corresponding to $\Delta\rho_p \sim 0$, $\Delta\rho_c \sim 0$ and five intermediate values close to $\Delta\rho_c = 0$. Changes in the scattering profile with variations in $\phi_h/\phi_d$ are shown in Figure 1 for a silica volume fraction sample of $\phi_c = 0.1$ in a solution containing $R_p = 0.4$ PEG where only the D/H ratio is varied. At low D–EtOH concentrations ($\phi_d/(\phi_h + \phi_d) \sim 0.15$), the scattering is dominated by the particles, which is proportional to $P_c(q)S_{cc}(q)$. As the D–EtOH concentration increases, substantial qualitative changes are observed in the scattering profile. To extract each $S_{ij}(q)$ from the above equation, we first determine $\Delta\rho_c$ and $\Delta\rho_p$ from the known $\rho_c$ and $\rho_s$ at each D/H–EtOH ratio using the information in Figure S2. Where $\Delta\rho_c = 0$, the above equation is simplified as $I(q) \sim A\Delta\rho_c{}^2P_c(q)S_{cc}(q)$. $S_{cc}(q)$ is obtained by dividing the scattering intensity from the concentrated particle suspension by its dilute limit analog at the same $R_p$. After the step, only $S_{pp}(q)$ and $S_{pc}(q)$ remain unknown. At each scattering vector, one then has five experimental data points at points according to the D/H–EtOH ratios and two unknowns allowing us to minimize uncertainty in the two unknowns at each $q$. We solve $S_{pp}(q)$ and $S_{pc}(q)$ using multiple linear regression fitting method.

**Fourier Transform Infrared Spectroscopy (FT−IR)**

The FT−IR spectra were collected with a VERTEX 80v (Bruker) in the range of $800-4000^{-1}$ cm.

**Simulation Method**

All-atom molecular dynamics (MD) simulations were performed to investigate the behavior of polyethylene glycol (PEG) chains with different molecular weights. While experimental PEG chains had molecular weights of 0.4 k and 10 k, simulating such long chains was computationally prohibitive. Therefore, shorter PEG chains with molecular weights of 0.2 k (N=5) and 1.1 k (N=25) were used as proxies. The simulation system comprised a silica slab periodic in the xy-plane, PEG chains, and ethanol. Four simulations were conducted by varying the weight percentage (wt%) and mole number of PEG with two different lengths and ethanol in the system (Table S1).

The simulations were carried out using GROMACS 2020.3,[37-39] with the CHARMM36 force field (July 2022),[40,41] supplemented by the CGenFF version 4.6[42-46] for parameterization. The NP$_z$T ensemble was employed to equilibrate the system, as it allows for pressure coupling along the z-axis while maintaining the xy-plane dimensions fixed. Since the silica slab is periodic in the xy-plane, it represents an infinite slab in these directions, and the pressure coupling is applied only along the z-axis to accommodate density fluctuations in the bulk PEG/ethanol mixture and ensure proper equilibration at the interface. This approach is commonly used for slab-like systems to stabilize interfacial properties without distorting the periodic slab geometry. The simulation box dimensions were initially set to 8.35 x 8.51 x 12.5 nm. The density of the system, excluding the silica slab, was adjusted so that the combined PEG (0.2 k and 1.1 k) and ethanol molecules achieved a density of 1.0 g/cm$^3$ at the start of the simulation.

The system was first equilibrated in the NP$_z$T ensemble at 363 K and 1 bar for 10 ns to stabilize temperature and pressure. To achieve rapid equilibration, the velocity-rescaling thermostat[47] was used to control the temperature, while the Berendsen barostat[48] was employed for pressure coupling. As the silica slab was periodic in the xy-plane, the pressure coupling was applied only along the z-axis. Following the initial equilibration, a more accurate thermodynamic ensemble was achieved by conducting an additional 10 ns NP$_z$T simulation using the Nosé–Hoover thermostat[49-51] and Parrinello–Rahman barostat[52] under the same conditions. Subsequently, production runs were performed in the NVT ensemble for 20 ns at 363 K, using the Nosé–Hoover thermostat. The final 2 ns of the production runs were analyzed to extract equilibrium properties. As shown in Figure S3, the total energy fluctuation during the 20 ns NVT simulation was minimal, confirming that the system was sufficiently equilibrated within the last 2 ns.

All bonds involving hydrogen atoms were constrained using the LINCS algorithm,[53] allowing a time step of 2 fs. Electrostatic interactions were treated using the Particle Mesh Ewald (PME) method[54] with a real-space cutoff of 1.2 nm. The PME interpolation order was set to 4, and the Fourier grid spacing was set to 0.16 nm.

**Density measurement**

The densities of polymers and PNCs were measured with a gas displacement pycnometer (AccuPyc II). Each sample was measured at least 9 times, and the average density was used.

## RESULTS AND DISCUSSION

We first composed selective adsorption with only long (or short) polymers and then mixed them with short (or long) polymers finally producing bidisperse PNCs. Figure S1 illustrates the experimental procedure: two bidisperse PNCs were prepared where either low (0.4k) or high (10k) MW PEG was preferentially adsorbed onto NPs (Figure S1 (a) and (b)). Then two PNCs were annealed for a sufficient time (7 days) to allow possible chain exchange ((c) and (d)). We compared the four PNCs: "0.4k first PNC" and "10k first PNC" both before and after the annealing. Unless otherwise specified, the volume ratio of PEG 0.4k to 10k is maintained at 1:1, and particle volume fraction ($\phi_c$) is fixed at 0.1. Then, the MW-dependent adsorption layers were compared for all four PNCs (Figure S1) where all chemical compositions are identical.

While the chemical compositions of four PNCs are identical, the particle surface of (a) and (b) is selectively enriched with 10k and 0.4k polymers, respectively and these selective adsorption layers can be altered in (c) and (d) after the following annealing process.

To directly observe the changes in the adsorbed polymer layers, contrast-matched small-angle neutron scattering (SANS) experiments were conducted. To avoid potential energetic biases in adsorption behavior between hydrogenous and deuterium-labeled PEG chains on silica particles,[55] we used highly concentrated solutions and contrast-matched the scattering length density (SLD) of the silica with deuterated ethanol (D–EtOH) polymer solutions, as previously reported[56,57] and shown in Figure S2.

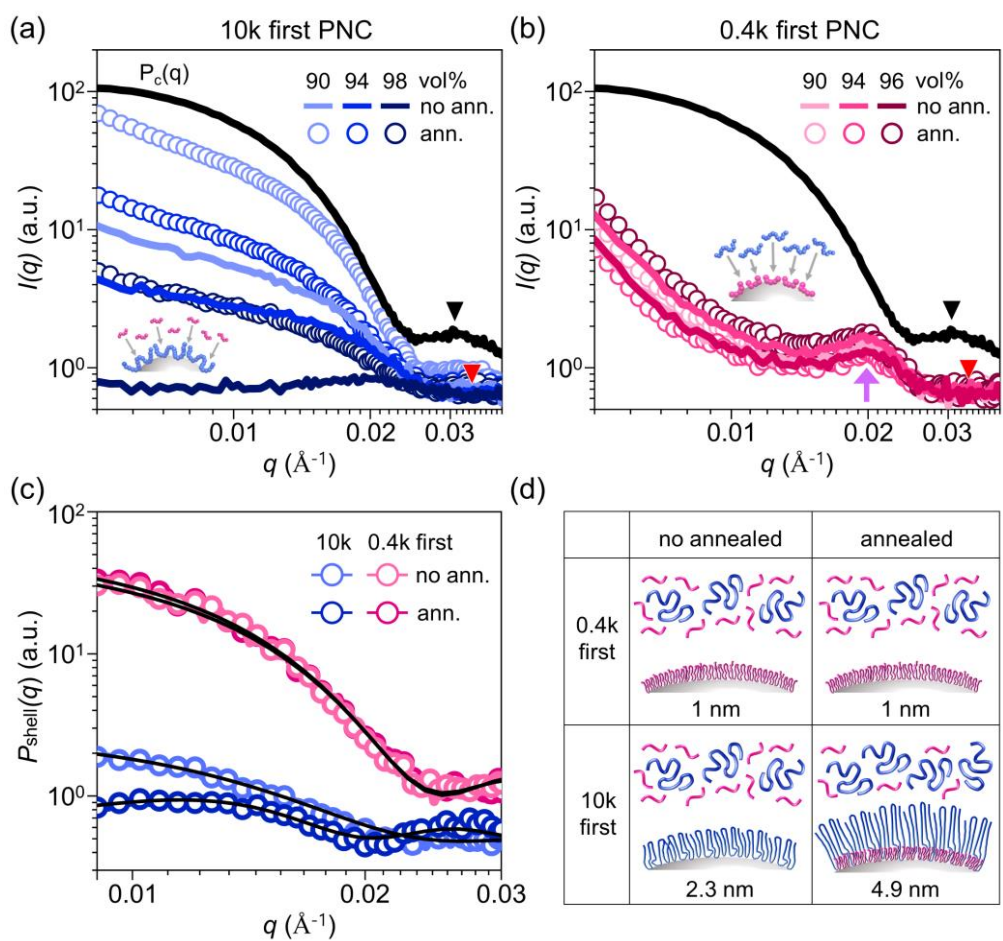

**Figure 1.** Contrast matched SANS intensity profiles for (a) 10k first and (b) 0.4k first PNCs with D-EtOH concentration before and after thermal annealing. (c) Polymer shell form factor, $P_{shell}(q)$, calculated from the corresponding SANS intensities. The black lines represent the form factor fit. (d) Schematic illustration for bulk and adsorbed polymers in 0.4k and 10k first PNCs before and after thermal annealing.

Figure. 1(a) and 1(b) show the changes in scattered intensity, $I(q)$ for 0.4k and 10k first PNCs as the SLD of the medium varies. For the 10k first, a progressive decrease in intensity is observed as the SLD of the medium matches that of silica (from "a" to "e" in Figure S2); the scattering from the silica significantly reduces and the scattering from the adsorbed layer of polymers arises. The oscillation peak at $q \sim 0.03$ Å$^{-1}$ from the form factor of silica NPs (black triangle) gradually diminishes and becomes almost undetectable at the matched condition. This trend indicates successful contrast matching with the silica particles, resulting in the polymer-polymer correlations becoming the dominant contributor to the scattering intensity. Notably, a new peak emerged at $q=0.035$ Å$^{-1}$ (red triangle), attributed to the form factor of the adsorbed polymer.

Similarly, the 0.4k first PNC shows a significant reduction in intensity near matched conditions with the absence of the silica signature peak, suggesting that scattering arises from the adsorbed layer. We found that the scattered intensity from the adsorbed polymer layers is distinguished from that of silica particles and exhibits a prominent peak at $q \sim 0.02$ Å$^{-1}$. Considering that the characteristic length of the bulk PEG appears at a much higher $q$ ($0.15 \sim 0.76$ Å$^{-1}$ based on their radius of gyration, $R_g$), the newly emerged peak at $q \sim 0.02$ Å$^{-1}$ (purple arrow) is attributed to spatial correlations between adsorbed polymers on a much large scale, exhibiting non-bulk-like properties.[56]

To obtain the structural information about the adsorbed layers in more detail, we extracted the polymer-polymer structure factor, $S_{pp}(q)$, from the series of $I(q)$ near the matched conditions by solving the simultaneous equation as shown in Figure S4. The $S_{pp}(q)$ provided consistent information with the matched condition of $I(q)$.

We then examined the changes in adsorbed layers of PNCs after the thermal annealing. During annealing, the selectively adsorbed polymer layer is expected to undergo modification, as competitive adsorption can lead to the replacement of previously adsorbed layers. Figure. 1(a) and 1(b) shows the SANS intensity profiles of the 10k and 0.4k first PNCs after annealing. For the 0.4k first PNC (Figure 1(b)), no noticeable difference in intensity is observed before and after annealing. This indicates the initially adsorbed short chains exhibit resistance to the challenges posed by the longer polymer.

However, intriguingly, 10k first PNC (Figure 1(a)) exhibits significant changes in intensity profiles following annealing, suggesting that the 0.4k PEG in bulk influences the adsorbed layer. This indicates that although PEG 10k initially adsorbs onto the particle surface, chain conformation at the interface changes due to the penetration of PEG 0.4k from the bulk. The broad peak observed in the annealed 10k first PNC implies the formation of a thicker polymer layer.

To quantitatively analyze and solely obtain information about the adsorbed polymer shell, we extracted the polymer shell form factor, $P_{shell}(q)$, and fitted it with a core-shell model to estimate the shell thickness as shown in Figure 1(c). Detailed procedures for obtaining $P_{shell}(q)$ are provided in Supplementary Information.

Figure 1(d) summarizes the interfacial changes and the variation of the layer thickness of adsorbed polymers. The 0.4k first PNC forms an adsorbed layer of ~1 nm thickness, corresponding to the $R_g$ of pre-adsorbed PEG 0.4k, which remains the same as $P_{shell}(q)$ shows no notable changes after annealing.

In contrast, the $P_{shell}(q)$ of 10k first PNC without annealing exhibited the peak at a lower $q$ value compared to the 0.4k first PNC, indicating the formation of a thicker adsorption layer (~2.3 nm), attributed to the longer chain length of pre-adsorbed PEG 10k. Following annealing, a peak further shifts to the left, implying the formation of an even thicker polymer shell (~4.9 nm) resulting from changes in the adsorbed polymers and their conformations.

Taken together, when short chains are adsorbed first, there is no change in the polymer shell layer. However, when long chains are initially adsorbed, the layer thickness increases after annealing, presumably due to the migration of short-chain polymers from the bulk to the interface. These contrasting behaviors of adsorbed layers provide compelling evidence for the preferential adsorption of shorter chains.

Theoretically, the preferential adsorption of shorter polymers on particle surfaces is often attributed to entropy-driven mechanisms; shorter chains experience a smaller entropy penalty upon adsorption when there is no energetic preference for polymer adsorption.[19-21] However, even though the short and long PEG chains are chemically identical, an energetic preference may arise.

PEG can form hydrogen bonds with the silanol groups (Si–OH) on silica NPs through its ether (C–O–C) bonds along the backbone as shown in Figure S5.[58] The hydroxyl (O–H) groups at the chain end, can form more stable hydrogen bonds with silanol groups. Since PEG 0.4k has significantly more end groups per unit volume than PEG 10k, it has a greater probability of forming hydrogen bonds. Thus, there can be additional enthalpic contribution to the preferential adsorption of shorter chains.

To elucidate the driving force behind the preferential adsorption, we examined the interactions between PEG and silica using Fourier transform infrared (FT–IR) spectroscopy. Figure 2 shows the FT–IR results for silica particles, neat polymers, and PNCs for both 0.4k and 10k PEG.

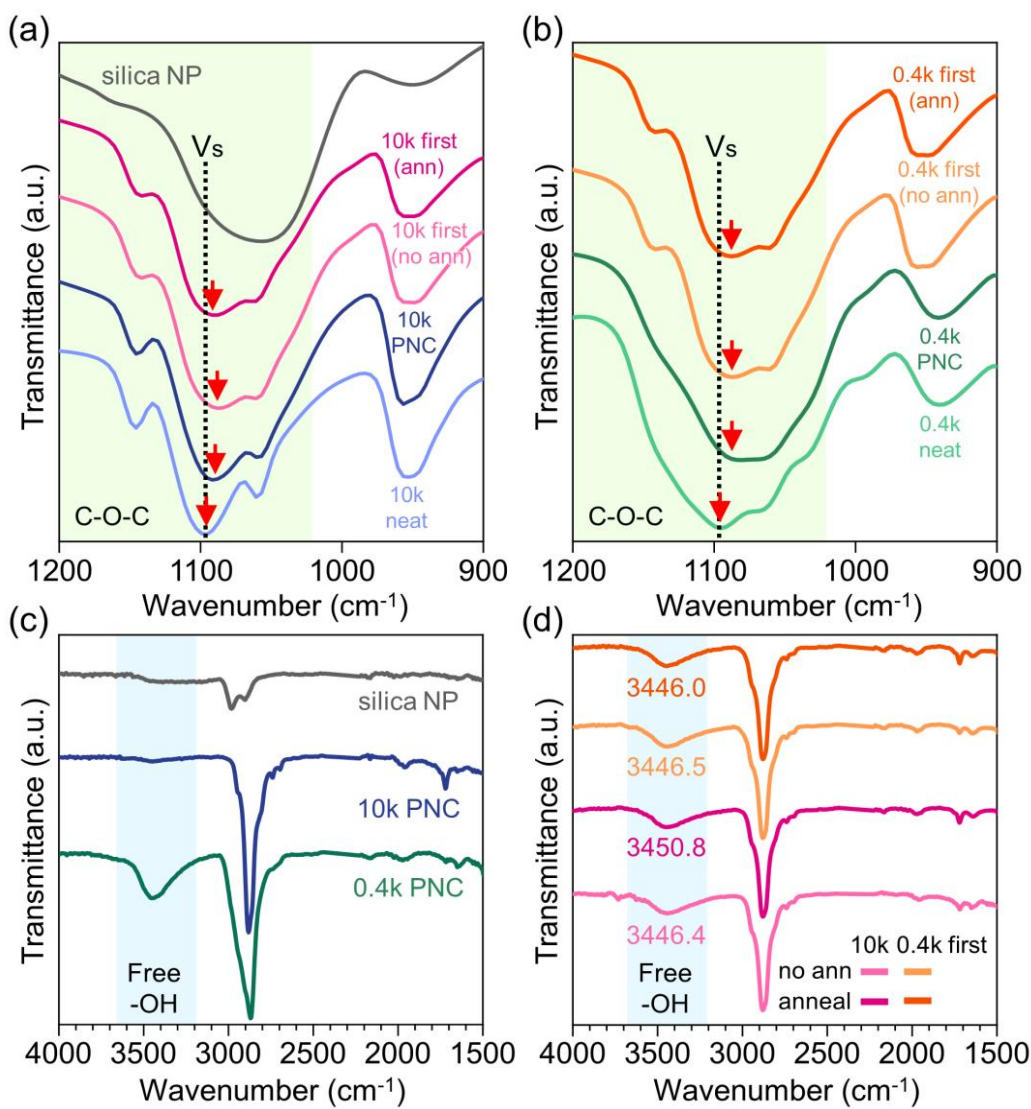

**Figure 2.** FT-IR spectra (1200 to 900 cm⁻¹) for (a) silica NP, 10k PNC (without 0.4k PEG), and 10k first PNCs, and (b) 0.4k PNC (without 10k PEG) and 0.4k first PNC. FT-IR spectra (4000 to 1500 cm⁻¹) for (c) silica NP, 10k PNC, and 0.4k PNC, and (d) 0.4k and 10k first PNCs before and after thermal annealing.

In Figure 2(a), the peak at ~1095 cm⁻¹ for neat PEG 10k is attributed to the stretching vibration ($V_s$) of the C–O–C bonds in the PEG backbone. Upon interacting with silica particles (10k only PNC), this peak red shifts to ~1086 cm⁻¹, indicating a decrease in the electron density of the C–O–C bonds due to the formation of hydrogen bonding with the silanol groups.[58] Unfortunately, further analysis of the silanol group is challenging due to the overlapping peaks of the Si-OH group and the C–C bond in PEG at ~950 cm$^{-1}$.[59]

To examine the changes in the adsorption behavior of bidisperse PNCs, Figure 2(a) also shows the FT–IR spectra of 10k first PNCs before and after thermal annealing. Before annealing, the particle surface is predominantly covered with 10k chains, resulting in a C–O–C peak similar to that of 10k-only PNC. However, after annealing, an intriguing blue shift of the peak is observed from 1087 to 1091 cm$^{-1}$, resembling the spectrum of neat PEG 10k. Since the red shifts occur when the C–O–C bond forms hydrogen bonds with silanol groups, the observed blue shift suggests the dissociation of pre-existing hydrogen bonds between PEG 10k (C–O–C) and the silanol groups.

Figure 2(b) demonstrates that PEG 0.4k also exhibits a red shift of the C–O–C peak when interacting with silica. Then, the identical peak position is found for the 0.4k only PNC and 0.4k first PNC suggesting that the silica surface is predominantly adsorbed with PEG 0.4k chains. Even after annealing, the peak at 1087 cm$^{-1}$ remains unchanged. This invariable peak position indicates that the interaction between the 0.4k chains and the silica surface is maintained during the annealing process and there is no additional adsorption from 10k chains.

To further investigate this disruption in hydrogen bonding observed in Figure 2(a), we examined the free O–H peak. In Figure 2(c), the peaks at ~3440 cm$^{-1}$, attributed to the intermolecular hydrogen bonding of the terminal O–H groups, are observed only when PEG 0.4k is present. This peak is not observed in PEG 10k due to the significantly fewer end groups.

In Figure 2(d), we note that the free O–H peak is observed in all PNCs from the end O-H groups in PEG 0.4k. For quantitative analysis, the peak position was determined by fitting the data using Gaussian and linear functions. While thermal annealing has no effect on this peak in the 0.4k first PNC, the 10k first PNC displays a blue shift from 3446.4 to 3450.8 cm$^{-1}$, implying the relocation of the free O-H peak from the bulk to the interface.

These findings indicate that the initially adsorbed PEG 0.4k (shorter chain) maintains its interactions with the silica surface, resisting displacement by the bulk PEG 10k (longer chain), even after thermal annealing. In contrast, in the 10k first PNC, the initially adsorbed PEG 10k are displaced with annealing, allowing the bulk PEG 0.4k to migrate to the interface and modify the adsorption layer.

To investigate the competitive adsorption between long and short PEG chains, MD simulations were conducted with PEG chains of varying lengths at the same weight fraction. Figure 3(a) presents MD snapshots of a system containing silica surface, PEG 5-mer, and 25-mer at the same wt% in an ethanol solvent, both at the initial state and after 40 ns. The snapshots reveal that the shorter PEG exhibits significantly higher adsorption onto the silica surface compared to the longer PEG. Figures 3(b) and S6 present the radial distribution function (RDF), g(r), calculated for oppositely charged atoms between the silica surface and PEG, which exhibit attractive interactions. This analysis reveals the atomic interactions that most strongly influence PEG adsorption. In particular, the interaction between the hydrogen atoms on the silica surface and the oxygen atoms in PEG exhibits a pronounced peak at approximately 2 Å, compared to those of other atomic pairs. In general, RDF peaks below 3.5 Å are attributed to electrostatic attractions, and if the interacting atoms are capable of forming hydrogen bonds, they are considered as such. This result indicates that hydrogen bonding between the silica surface's hydrogen atoms and PEG's oxygen atoms plays a critical role in PEG adsorption.

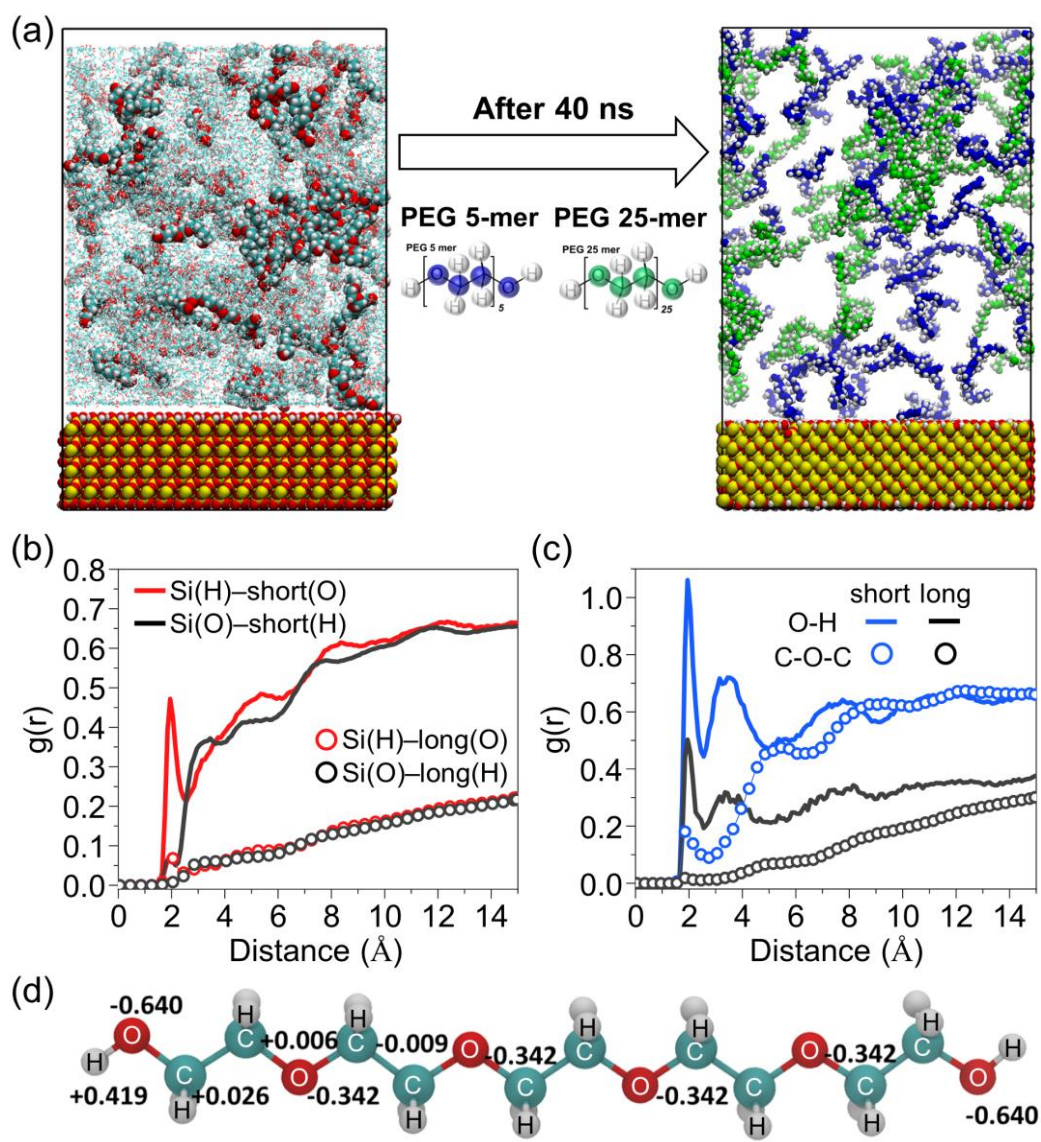

**Figure 3.** (a) MD snapshots of a system with a silica surface, PEG (5-mer and 25-mer) in ethanol, shown at initial state (left) and after 40 ns (right). (b) The radial distribution function, g(r), calculated for oppositely charged atoms between the silica and PEG molecules. (c) The radial distribution function, g(r), between hydrogen atoms on the silica and oxygen atoms in the C-O-C or O-H groups. (d) Partial charges of atoms assigned by the simulation.

To identify which oxygen-containing functional groups in PEG most strongly participate in hydrogen bonding with the hydrogens of silanol groups, we examined the RDF between silica surface hydrogens and oxygens of the C-O-C backbone and terminal O-H groups in PEG. Figure 3(c) shows a pronounced peak between O-H oxygens and silanol hydrogens, confirming that terminal O-H groups primarily form hydrogen bonds with silanol groups. This observation is consistent with the partial charges assigned in the simulation (Figure 3(d)), where the oxygen of the C–O–H group has a significantly larger negative charge (–0.640) than that of the C–O–C oxygen (–0.342). Furthermore, the higher RDF peak for shorter chains compared to longer ones suggests that the greater number of terminal O–H groups in shorter PEG chains leads to relatively stronger interactions with the silica particles.

Considering that the difference in the number of repeat units ($N$) used in the simulation (5-fold) is much smaller than the actual difference between PEG 0.4k ($N$=9) and PEG 10k ($N$=227), it can be expected that shorter PEG chains are significantly more likely to adsorb in the experimental system. These findings suggest that the preferential adsorption of shorter PEG chains arises largely due to the strong enthalpic contributions from differences in hydrogen bonding between the functional groups.

This preference for shorter chains is attributed to the substantially higher number of O–H groups per unit volume in PEG 0.4k than 10k. Therefore, when the number of moles (i.e., the number of end groups) of PEG 0.4k and 10k are identical, the energetic interaction preference of silanol for the O–H groups with chain lengths can be minimized.

Figure 4(a) shows the FT–IR spectra of the 10k first PNCs, prepared with varying molar ratios ($\alpha_{0.4k}$) of PEG 0.4k to PEG 10k. When the number of O–H groups in PEG 0.4k is lower than that in PEG 10k ($\alpha_{0.4k}$ < 1), no shift in the C–O–C peak is observed with thermal annealing. However, when the number of O–H groups in PEG 0.4k exceeds that in PEG 10k ($\alpha_{0.4k} \geq 1$), the C–O–C peaks blue-shift from ~1087 to ~1090 cm$^{-1}$. This shift indicates that preferential adsorption occurs due to the replacement of hydrogen bonds from the C–O–C groups of PEG 10k with those involving the terminal O–H groups of PEG 0.4k.

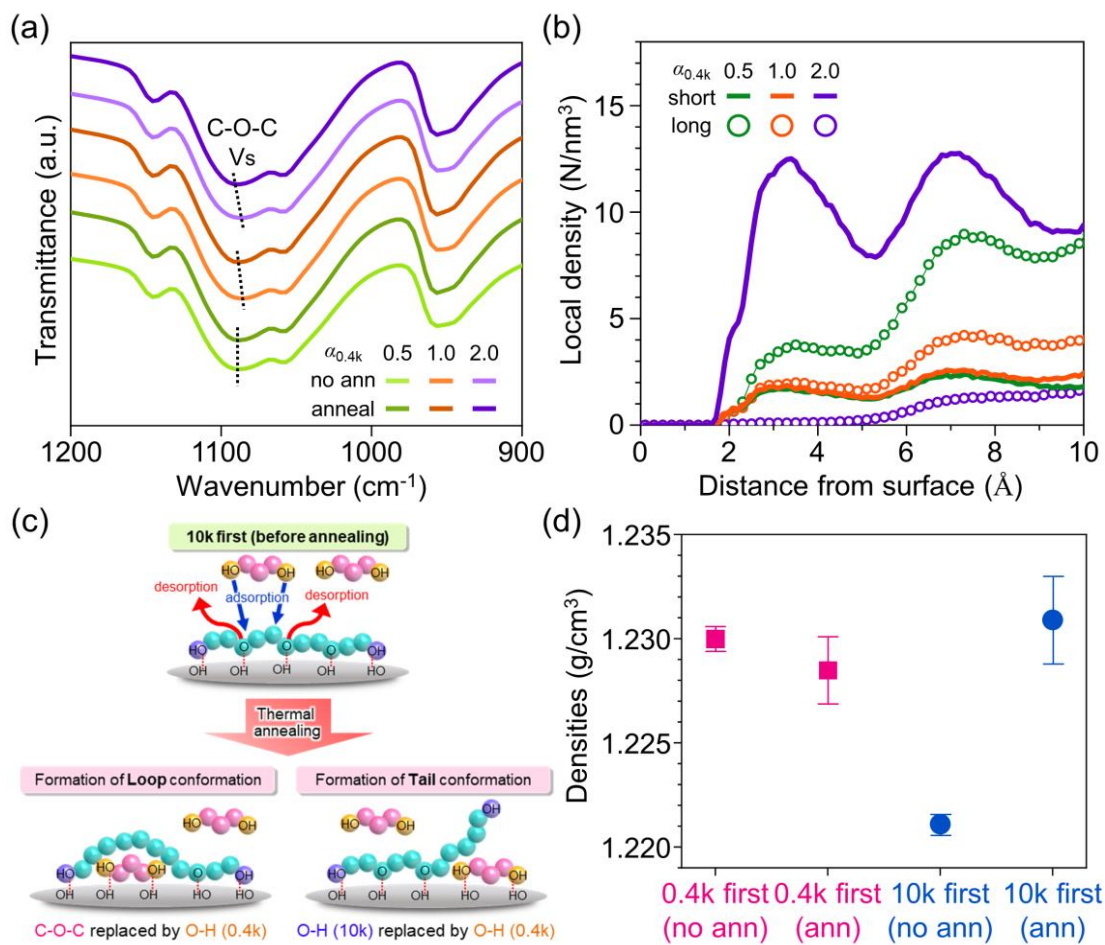

**Figure 4.** (a) FT-IR spectra and (b) local density profiles for 10k first PNC with varying mol ratio of PEG 0.4k and 10k. (c) schematic illustration for preferential adsorption of shorter chain in 10k first PNC. (d) Measured density of 0.4k and 10k first PNCs before and after thermal annealing.

This result is further supported by MD simulation observations (Figure 4(b)) showing that when the mol ratio of short-chain PEG is low, the local density of long chains is higher. However, as the molar ratio of short-chain PEG increases, the local density of the short chains increases more significantly. This confirms that the preferential adsorption of shorter chains is an enthalpically driven process.

Figure 4(c) illustrates the competitive adsorption behavior in the 10k first PNC. Initially, the silanol groups predominantly form hydrogen bonds with the C–O–C bonds of PEG 10k. However, as indicated by the changes in FT–IR spectral, these hydrogen bonds are gradually replaced by new hydrogen bonds involving the free O–H groups of the added 0.4k. Due to the high energetic barrier for whole chain desorption (>100 kT),[3] this replacement occurs through the desorption of segments of PEG 10k rather than the entire chain. In this case, a "loop" conformation is likely to form. Alternatively, the end O–H group of PEG 10k may also be replaced by the end O–H group of PEG 0.4k, leading to the detachment of PEG 10k segments near the end group and the formation of a "tail" conformation.

The adsorption of 0.4k PEG between 10k PEG chains can result in a denser interfacial layer. Genix et al. reported that as the MW increases, chain packing frustration occurs in the interfacial region, leading to lower density.[60,61] As shown in Figure 4(d), before annealing, the 10k first PNC exhibited lower density than the 0.4k first PNC. However, after annealing, unlike the 0.4k first PNC where there was no change in density, the 10k first PNC showed a significant increase in density. The fact that its density became similar to that of the 0.4k first PNC further supports the result of preferential adsorption of the 0.4k. The result indicates that a densely packed interfacial layer is formed through the exchange of long PEG segments and short PEG chains, promoting the elongation of the long chains and increasing the thickness of the SIL, as confirmed by SANS results.

The stretching of the SIL induces an entropy loss due to the reduced configurational freedom of the elongated chains. However, this entropy penalty can be compensated by the enthalpy gained from replacing the hydrogen bonds between the C–O–C groups of PEG 10k and the Si–OH groups with the stronger hydrogen bonds involving the terminal O–H groups of PEG 0.4k.

## CONCLUSION

Our comprehensive investigation, integrating both experimental observations and molecular dynamics simulations, demonstrates that shorter polymer chains preferentially adsorb onto surface in bidisperse polymer melts. This preferential adsorption arises from enthalpic contributions due to differences in the number of functional groups—specifically, the greater number of terminal hydroxyl groups in shorter chains—which enhance hydrogen bonding interactions with the surface, despite the polymers having identical chemical structures. This behavior leads to a densely packed interfacial polymer layer that induces stretching of the pre-adsorbed longer chain conformations, significantly influencing the interfacial properties.

These results highlight the critical impact of molecular weight on polymer adsorption behavior and underscore the importance of both chain length and end-group functionality in determining interfacial characteristics. Understanding these mechanisms provides valuable insights for tailoring interfacial interactions in polymer nanocomposites and offers a strategic pathway for designing advanced materials with optimized performance. Future work will focus on confirming how changes in the adsorbed layer affect macroscopic properties and particle dispersion structures within the composites, further enhancing our ability to engineer materials with specific functionalities through precise control of interfacial phenomena.

# Supporting information

Supporting information: Schematic diagram for the preparation of pre-adsorbed PNC samples, Scattering length density and contrast matching method for SANS, Total energy for MD simulation systems, Compositions for MD simulation systems. Partial structure factor for polymer-polymer correlation, Calculated radial distribution function.

# Author Information

**Corresponding Authors**

So Youn Kim – Department of Chemical and Biological Engineering, Institute of Chemical Processes, Seoul National University, Seoul 08826, Republic of Korea; Email: soyounkim@snu.ac.kr

YongJoo Kim – Department of Materials Science and Engineering, Korea University; Seoul, 02841, Republic of Korea; Email: cjyjee@korea.ac.kr

**Authors**

Tae Yeon Kong – Department of Chemical and Biological Engineering, Institute of Chemical Processes, Seoul National University, Seoul 08826, Republic of Korea

WooJin Kim – Department of Materials Science and Engineering, Kookmin University; Seoul 02707, Republic of Korea

**Notes**

The authors declare no competing financial interest.

# Acknowledgments

This work was supported by the National Research Foundation of Korea (NRF) grant funded by the Korea Government (MSIT) (NRF-2021R1A2C2007339 and NRF-

# Supporting Information

Competitive Adsorption in Polymer Nanocomposites: The Molecular Weight and End-Group Effect Revealed by SANS and MD Simulations

Tae Yeon Kong[1], WooJin Kim[2], YongJoo Kim[3,*], So Youn Kim[1,*]

[1]Department of Chemical and Biological Engineering, Institute of Chemical Processes, Seoul National University; Seoul, 08826, Republic of Korea

[2]Department of Materials Science and Engineering, Kookmin University; Seoul 02707, Republic of Korea

[3]Department of Materials Science and Engineering, Korea University; Seoul, 02841, Republic of Korea

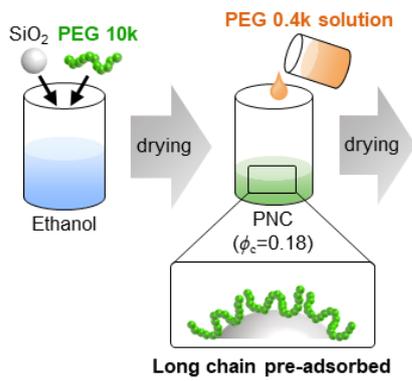

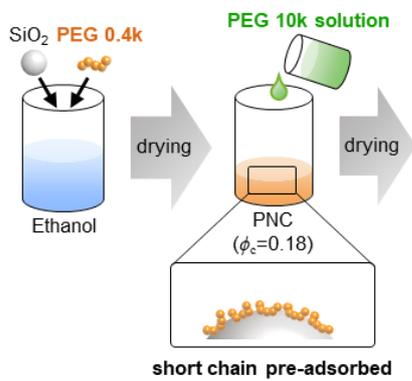

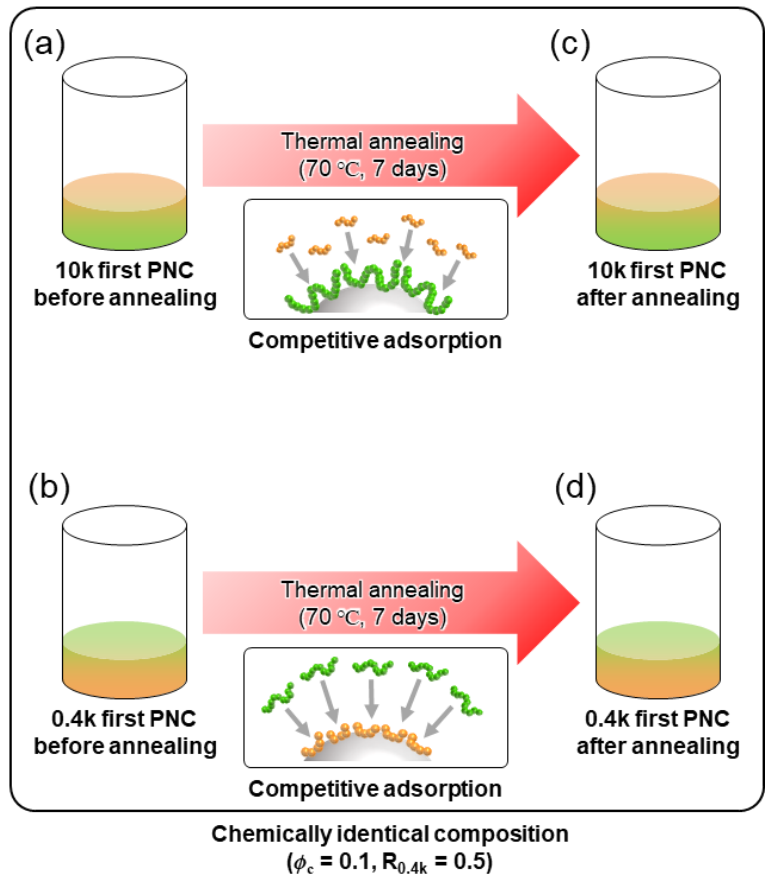

**Figure S1.** Sample preparation process for 10k and 0.4k first PNCs before and after thermal annealing. Chemical compositions for all PNCs are identical ($R_{short}$=0.5 and $\phi_c$=0.1).

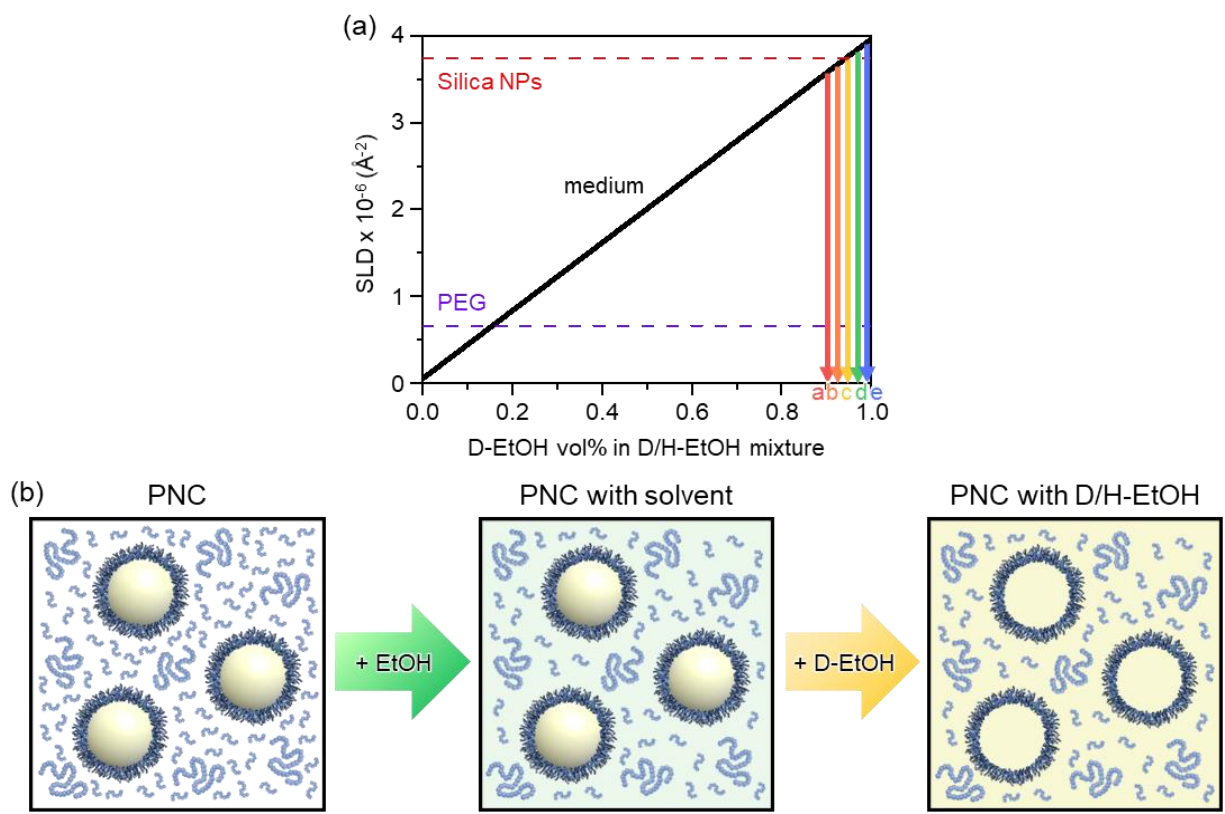

**Figure S2.** (a) Scattering length density (SLD) as a function of D-EtOH vol% in D/H-EtOH mixture. (b) Contrast matching method to selectively observe the adsorbed polymer layers.

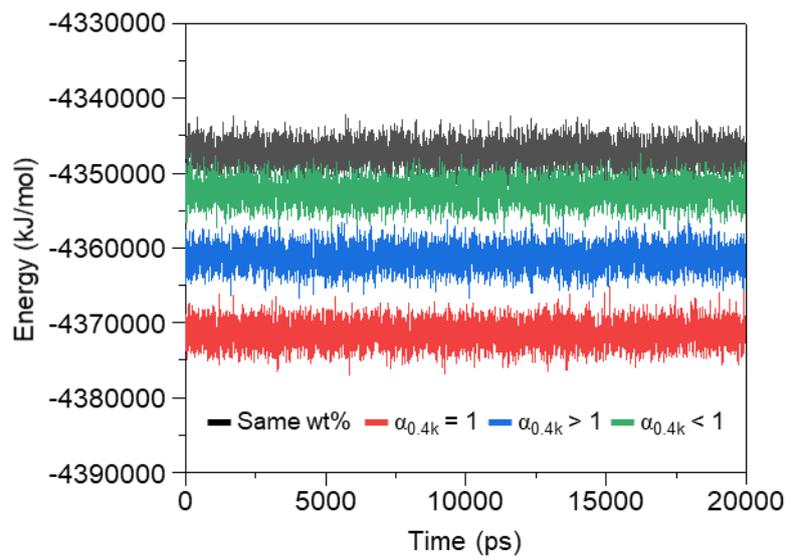

**Figure S3.** The total energy of the four MD simulation systems during the 20 ns of NVT production simulation runs.

**Table S1.** Compositions for four different MD simulation systems.

| System | PEG 5-mer (number) | PEG 25-mer (number) | PEG 5-mer (wt%) | PEG 25-mer (wt%) | Ethanol (number) | Ethanol (wt%) |
|---|---|---|---|---|---|---|
| 1. same wt% | 86 | 18 | 6.0 | 6.0 | 6570 | 88.0 |
| 2. $\alpha_{0.4k} = 1$ | 18 | 18 | 1.3 | 6.0 | 6924 | 92.7 |
| 3. $\alpha_{0.4k} > 1$ | 86 | 9 | 6.0 | 3.0 | 6794 | 91.0 |
| 4. $\alpha_{0.4k} < 1$ | 14 | 31 | 1.0 | 10.0 | 6645 | 89.0 |

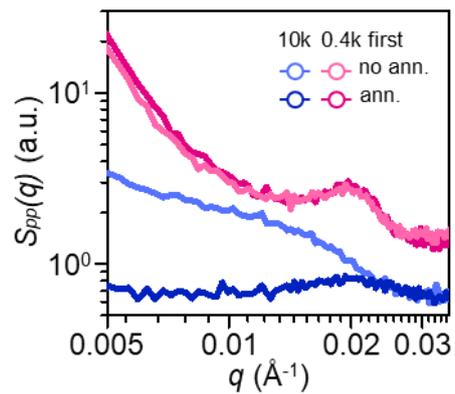

**Figure S4.** Partial structure factor for polymer-polymer correlation, $S_{pp}(q)$, for 10k and 0.4k first PNCs before and after thermal annealing.

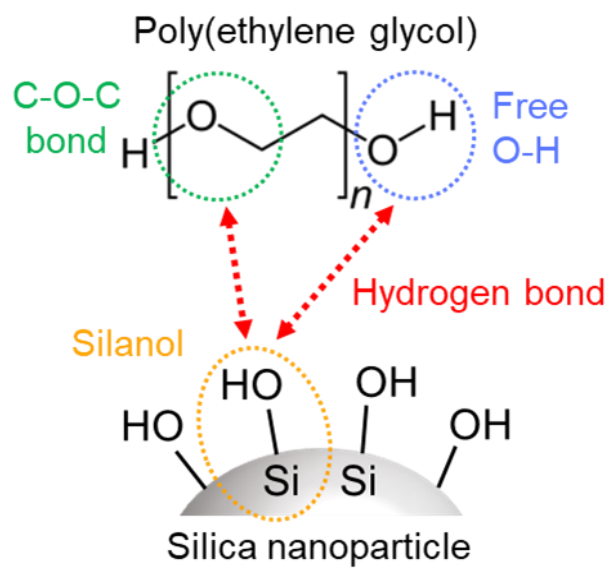

**Figure S5**. Hydrogen bonding between PEG and silica nanoparticle.

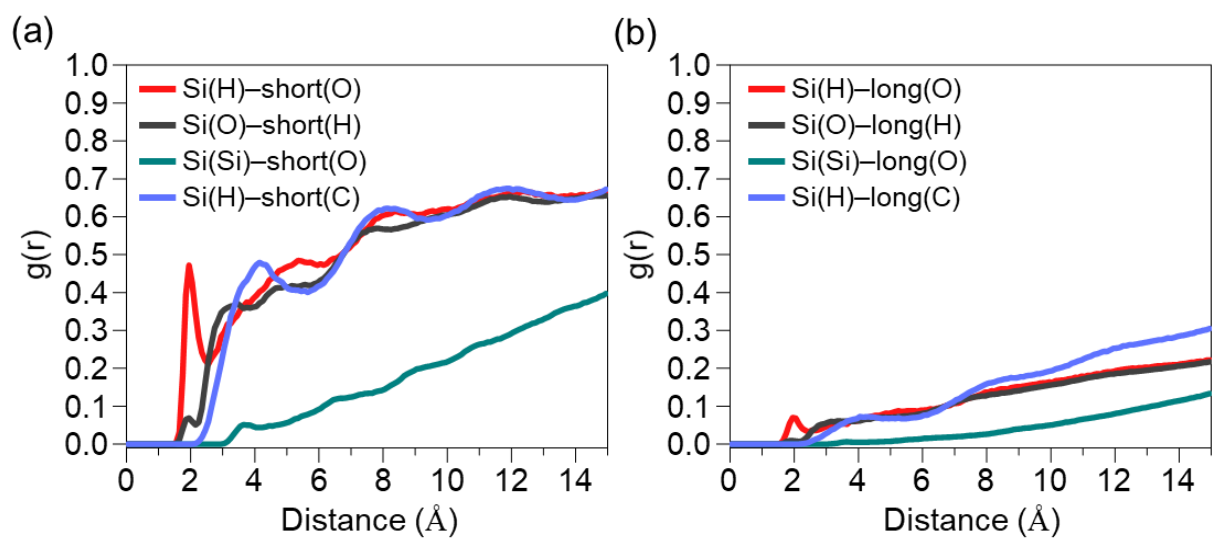

**Figure S6.** The radial distribution function, g(r), calculated for oppositely charged atoms between the silica and PEG molecules.